\documentclass[aps,prl,twocolumn,groupedaddress,showpacs]{revtex4}
\usepackage{graphicx}  % Include figure files

\begin{document}

\title{Magnetoresistance from quenching of spin quantum correlation in organic semiconductors}

\author{Wei Si}\affiliation{State Key Laboratory of Surface Physics and Department of Physics, Fudan University, Shanghai 200433, China}

\author{Yao Yao}\affiliation{State Key Laboratory of Surface Physics and Department of Physics, Fudan University, Shanghai 200433, China}

\author{Xiaoyuan Hou}\affiliation{State Key Laboratory of Surface Physics and Department of Physics, Fudan University, Shanghai 200433, China}

\author{Chang-Qin Wu}\email[Email: ] {cqw@fudan.edu.cn} \affiliation{State Key Laboratory of Surface Physics and Department of Physics, Fudan University, Shanghai 200433, China}

\date{\today}

\begin{abstract}
We present a theory of organic magnetoresistance (OMR) based on the quenching of the quantum correlation between the carrier's spin and its local environment when the incoherent hopping takes place. We prove that this process contributes a spin-dependent prefactor to the attempt-to-escape frequency in the hopping rate, with its value modulated by the magnetic field. The resulting OMR exhibits a positive Lorentzian saturation component and a negative small-field component, which are independent of model parameters. These behaviors, with their isotope effects, are in good agreement with experimental results.
\end{abstract}


\maketitle

\section{Introduction}

The organic magnetoresistance (OMR) has attracted much attention since its discovery \cite{discCPL, discNJP}, due to its unique features and potential applications in magnetic sensors. It is a sizable (up to 5\%) and robust effect under weak magnetic field (tens of milli-Tesla) and room temperature, which is observed in a wide range of amorphous organic semiconductors (OSC) with surprising generality. The OMR behavior can be fitted well by a Lorentzian ($B^2/(B^2+B_0^2)$) or a non-Lorentzian ($B^2/(|B|+B_0)^2$) lineshape. The sign of the OMR can be tuned by applied voltage \cite{KoopmansSign, GillinSign}, device structure \cite{HuSign, EpsteinSign} and temperature \cite{KoopmansSignT}. Typically, the electric current increases with magnetic field in bipolar device \cite{GillinAlq3, LiPulse}, while decreases in unipolar ones \cite{VardenyUltrasmall}. Recently, efforts have been taken to clarify the isotope effects of OMR, which leads to different conclusions in small-molecule \cite{GillinIsotope} and polymer \cite{VardenyIsotope} devices. Another important advance is made by Nguyen \textit{et al.} with the discovery of an ultrasmall-field component, which scales with the main component and takes an opposite sign \cite{VardenyUltrasmall}. This finding provides more clues on the underlying mechanism of the OMR.

The origin of the OMR is recognized to be the spin interactions in OSC, such as the hyperfine interaction \cite{WohlgHyperfine} and the spin-orbit coupling \cite{DrewSO}. Several microscopic processes have been proposed to be responsible, including the blocking of carriers by bipolarons \cite{BobbertBipolaron} and excitons \cite{GillinAlq3}, interfacial dissociation of excitons \cite{GillinComponent}, and electron-hole pair mediated processes \cite{VardenyUltrasmall, BobbertPair, YaoMobility}. Simulations by kinetic Monte Carlo method and stochastic Liouville equations have achieved satisfying comparison with many experimental results. However, it is realized that the observed OMR should be the net effect of multiple components \cite{VardenyBlend, GillinComponent} from the abundant electronic processes. A comprehensive understanding of the phenomena is still lacking despite much research efforts.

The OMR should stem from the interplay between the dynamics of the charge carriers and that of their spins in OSC. The relative Hamiltonian for the system can be separated into two parts for the charge and the spin. However, their energy scales are distinctly separated. The charge part contains, for example, the transition between different transport sites and the interaction with lattice vibrations. The second part consists of the spin interactions.  The dynamics of the former one commonly lie in the picoseconds time scale. In contrast, the spin interactions in OSC are in the $0.1\mu$eV regime and the related dynamics are coherent in the nanoseconds time scale \cite{DrewMuon}. Thus most existing theories treat the two parts separately. The charge part renders the incoherent hopping of charge carriers among different localized electronic states assisted by the lattice vibrations. This is distinct from the band transport of extended states in inorganic semiconductors. Phenomenological approach is the most effective tool till now to describe the hopping transport, due to the complexity of the lattice vibrations and the decoherence process. In these works, the transport sites are assumed to have random energies \cite{BasslerMC}. The hopping rate is given by either the Miller-Abrahams (MA) theory \cite{MillerA} or the Marcus theory \cite{Marcus}. For example, the MA formula for the hopping rate between site $i$ and site $j$ is
\begin{equation}
\nu_{ij}=\left\{
\begin{array}{ll}
\nu_e\exp[-(\epsilon_j-\epsilon_i)/k_BT], & \epsilon_j>\epsilon_i,  \\
\nu_e, & \epsilon_j \leq \epsilon_i,
\end{array}
\right.
\label{ma}
\end{equation}
where $\nu_e$ is the attempt-to-escape frequency; $\epsilon_j$ ($\epsilon_i$) is the energy of the electronic states on site $j$ ($i$); $k_B$ is the Boltzmann constant and $T$ is the temperature. To incorporate the spin dynamics into the above framework, several methods have been proposed, such as the semiclassical \cite{BobbertDiffusion} approach and the Franck-Condon-like one \cite{YuSO}. Yet in these approaches, the spin-related quantum coherence in the presence of the incoherent hopping is not specifically considered. They are among the essential features of the organic spintronics compared with the traditional ones and worth further study.

In this paper, we focus on the quantum correlation between the carrier's spin and the local environment of spin (LES) formed by the spin interactions. The spin quantum correlation is expected to be quenched by the incoherent hopping of charge carriers. The quenching results in a prefactor $\eta$ in the attempt-to-escape frequency $\nu_e$ of Eq. (\ref{ma}). Furthermore, the value of $\eta$ is determined by the degree of the quantum correlation. By this process, an external magnetic field could influence the hopping rate by modulating the quantum dynamics of the carrier's spin and the LES. As a result, the magnetic field alters the carrier's mobility, which leads to OMR. The paper is organized as follows: Section 2 describes the theory of the OMR from quenching of spin quantum correlations by incoherent hopping of carriers. In section 3, quantitative results from a hyperfine interaction model are shown. A brief conclusion is drawn in Section 4.

\section{Theory}

\begin{figure}
\includegraphics[angle=0]{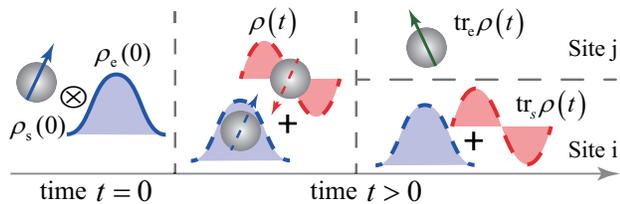}
\caption{Illustration of an incoherent hopping process with spin interactions. The carrier's spin is denoted by the arrow and the local environment of spin (LES) by the waveform. The LES can be nuclear spins of hydrogen atoms \textit{etc.}. At $t=0$, the quantum state of the composite system is separable; For the hopping attempt at time $t>0$, the state of the composite system before hopping is quantum-correlated caused by the spin interactions. However, the quantum correlation is quenched by the incoherent hopping to site $j$ and the final state of the composite system is separable again.}
\label{Fig1}
\end{figure}

First we present the origin of the prefactor $\eta$ in the attempt-to-escape frequency. The physical process is illustrated by a two-site model shown in Fig. \ref{Fig1}, with a carrier hopping from site $i$ to site $j$, which is the basic process of charge transport.
At time $t=0$, a charge carrier hops onto site $i$. During the carrier stays on site $i$, the interaction with the lattice vibrations will constantly decohere the carrier's dynamics and drive the carrier to make attempts to hop incoherently to site $j$. Each hopping attempt happens on the decoherence time scale of picoseconds \cite{YaoDecohere}. Simultaneously, the carrier's spin will interact with the local environment through spin interactions, such as the hyperfine interaction with nuclear spins of the hydrogen atoms. We term the environment as the local environment of spin (LES). The two subsystems of the carrier's spin ($\rho_{\rm s}$) and the LES ($\rho_{\rm e}$) together constitute the spin-related composite system ($\rho$) that is concerned in the following. At $t=0$, the quantum states of the two subsystems are independent, \textit{i.e.}, the density matrix of the composite system $\rho$ are separable as $\rho(0)=\rho_{\rm s}(0)\otimes\rho_{\rm e}(0)$. However, for the hopping attempt at time $t>0$, quantum correlation is generated between them. For the state of the composite system after the hopping, it is noted that the incoherent hopping acts as a local measurement process of the system and will also disturb the spin-related quantum coherence. In such a system, at least three types of coherence can be identified. They include the individual quantum coherence of the two subsystems, which can be reflected by the off-diagonal matrix elements of the reduced density matrix of the carrier's spin/LES $\rho_{\rm s/e}(t)={\rm tr}_{\rm e/s}\{\rho(t)\}$, where ${\rm tr}_{\rm e/s}$ denote the partial trace over the degrees of freedom of the LES/carrier's spin. They are expected to survive the hopping process as the spin interaction is weak. The third type is the quantum correlation between the two subsystems, which is dominated by a different set of off-diagonal density matrix elements \cite{YuESD}. This coherence would be a nonlocal one when the carrier is on site $j$ and therefore it should be quenched after the \textit{incoherent} hopping \cite{YaoESD}. To satisfy the above requirements, we use the the adiabatic elimination procedure \cite{AdiabaticE}. For the hopping attempt at time $t$, the initial state density matrix of the composite system is $\rho(t)$ and the final state one is $\rho_{\rm s}(t)\otimes\rho_{\rm e}(t)$. According to the Fermi golden rule, this process results in a prefactor $\eta(t)$ in the attempt-to-escape frequency $\nu_e$, which reads \cite{si}
\begin{equation}
\eta(t)=\mathrm{tr}\left\{\rho(t)[\rho_{\rm s}(t)\otimes\rho_{\rm e}(t)]\right\}.
\end{equation}
Further, the value of $\eta(t)$ is determined by the degree of quantum correlation between the carrier's spin and the LES, with larger correlation corresponding to smaller $\eta(t)$ \cite{si}. By this process, an external magnetic field could influence the hopping rate by modulating the quantum dynamics of the composite system. Finally, the phenomenological carrier mobility is magnetic-field dependent.

In the following, to focus on the consequence of $\eta(t)$, we assume the ensemble average concerning $\eta(t)$ are statistically independent with that over other quantities, such as the site energies. A parameter $\nu_0$ is introduced to account for the ensemble-averaged value of those parts in Eq. (\ref{ma}). The hopping rate is written as $\nu_0\eta(t)$. We take a hyperfine interaction model to present the theory quantitatively, in which the nuclear spins of the hydrogen atoms act as the LES. The Hamiltonian takes the form \cite{VardenyIsotope}
\begin{equation}
H_s = \sum_\alpha J_\alpha\mathbf{I}_\alpha\cdot\mathbf{s}+g\mu_BBs_z,
\label{hami}
\end{equation}
where we have set $\hbar=1$; $\alpha$ is the label for nuclear spins in the LES, with $J_\alpha$ the corresponding coupling strength and $\mathbf{I}_\alpha$ the nuclear spin operators; ${\bf s}$ are the carrier's spin operators; $g$ is the Land\'{e} factor which is taken to be $2$; $\mu_B$ is the Bohr magneton. The evolution of the composite system under Eq. (\ref{hami}) results in a time-dependent $\eta(t)$.  For a time-independent quantity which reflects the OMR, we assume that the hopping attempts are Markovian. Suppose the probability that the carrier is still on site $i$ at time $t$ is $P(t)$, the master equation for $P(t)$ is ${\rm d}P(t)/{\rm d}t = -\nu_0\eta(t)P(t)$. The average waiting time $t_{\rm w}$ is
\begin{equation}
t_{\rm w}=\int_0^{\infty}\exp\left(-\nu_0\int_0^{t}\eta(\tau)\mathrm{d}\tau\right)\mathrm{d}t.
\label{tw}
\end{equation}
The inverse of the average waiting time, denoted as $\nu=1/t_{\rm w}$, serves as an effective hopping rate. If the above-mentioned quenching process is neglected with $\eta(t)=1$, the expression returns to the one used in kinetic Monte Carlo simulations as $\nu=1/t_{\rm w}=\nu_0$. Further, $\nu$ can be connected to the carrier mobility by the relation
\begin{equation}
\mu=\nu a/\xi E=\nu aL/\xi V,
\label{mobility}
\end{equation}
where $L$ is the thickness of the device; $a$ is the lattice constant; $\xi$ reflects the randomness of the hopping directions; $E$ is the average electric field in the device and $V$ is the external voltage. This relation serve as an approximation by neglecting the fluctuations present in the complete Monte Carlo simulations. From Eq. (\ref{mobility}), the magnetic-field effect of the hopping rate is reflected by that of the carrier mobility with
\begin{equation}
\frac{\mu(B)-\mu(0)}{\mu(0)} = \frac{\nu(B)-\nu(0)}{\nu(0)}\equiv \frac{\Delta\nu}{\nu}.
\end{equation}
Therefore in the following the magnetic-field dependence of $\nu$ is discussed for the OMR effect.

\section{Results}

\begin{figure}
\includegraphics[angle=0]{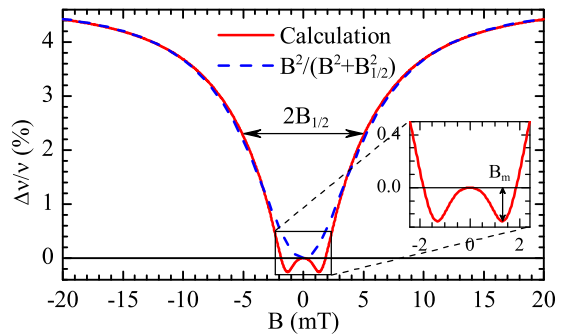}
\caption{The magnetic-field dependence of the hopping rate $\Delta\nu/\nu$ with $J=0.2\mu$eV and $\nu_0=3.5J$. The dashed line is the fit of the saturation component by the Lorentzian lineshape, with $B_{1/2}$ the half-width at half-maximum (HWHM). The inset shows the negative component of $\Delta\nu/\nu$ at small fields, where $B_{\mathrm{m}}$ is the magnetic field under which the hopping rate is minimum.}
\label{Fig2}
\end{figure}

\begin{figure}
\includegraphics[angle=0]{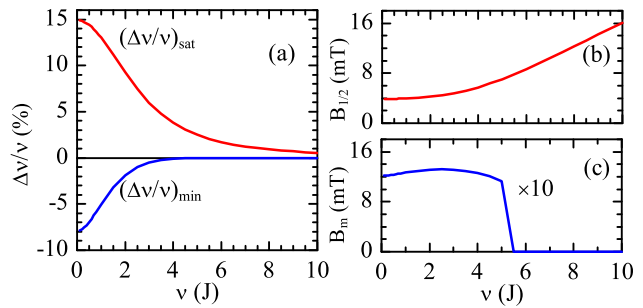}
\caption{The $\nu_0$-dependence of the magnetic-field dependence of the hopping rate $\Delta\nu/\nu$. The variation of the saturation value $(\Delta\nu/\nu)_{\mathrm{sat}}$ and the minimum value $(\Delta\nu/\nu)_{\mathrm{min}}$ with $\nu_0$ are shown in (a); the variation of the HWHM $B_{1/2}$ is shown in (b) and the variation of the magnetic field under which the hopping rate is minimum, $B_{\mathrm{m}}$, is shown in (c). The data of $B_{\mathrm{m}}$ has been multiplied by $10$ for clarity.}
\label{Fig3}
\end{figure}

We first consider the case with one nuclear spin-$\frac{1}{2}$ as the LES. The coupling constant is set to be $J=0.2\mu$eV and the parameter $\nu_0$ is taken to be $\nu_0=3.5J$. If we take a device thickness of $400$nm with an external applied voltage of $5$V and inter-site distance $0.5$nm, the $\nu_0$ value corresponds to a mobility of $\sim 4\times 10^{-4}\mathrm{cm}^2/(\mathrm{V}\cdot\mathrm{s})$ from Eq. (\ref{mobility}), which reflects the typical carrier mobility in the OSC, such as ${\rm Alq}_3$. The magnetic-field dependence of hopping rate $\nu$ is shown in Fig. \ref{Fig2}.
The final result is averaged over a set of initial states, with the carrier's spin taking any orientation and the LES taking any of its eigenstates in the magnetic field. One prominent feature is that two components are observed: a positive saturation one at large fields and a negative one at small fields, which is in correspondence with experimental observations \cite{VardenyUltrasmall, VardenyIsotope}. Besides, the saturation component is well fitted by a Lorentzian lineshape $B^2/(B^2+B_{1/2}^2)$. It should be emphasized that the two-component behavior is a robust and general outcome of our theory, which does not depend on the choice of parameters. For example, we show in Fig. \ref{Fig3}
the variation of this dependence with $\nu_0$. The characters of the dependence are reflected by four quantities, which are the saturation amplitude $(\Delta \nu/\nu)_{\mathrm{max}}$, the half-width at half-maximum (HWHM) $B_{1/2}$, the minimum value $(\Delta \nu/\nu)_{\mathrm{min}}$ and the corresponding magnetic field $B_{\mathrm{m}}$. With increasing $\nu_0$, the integrand of Eq. (\ref{tw}) decays faster and there is less time to build up the quantum correlation, so both $(\Delta \nu/\nu)_{\mathrm{sat}}$ and $(\Delta \nu/\nu)_{\mathrm{min}}$ decrease. The negative component vanishes for $\nu_0$ beyond about $5J$, as the oscillatory behavior of the quantum correlation plays a minor role in these cases. Furthermore, while $B_{\rm m}$ remains nearly unchanged, $B_{1/2}$ increases steadily with $\nu_0$, which takes the same trend of increasing $B_{1/2}$ with electric field observed in experiment \cite{VardenyIsotope}. These behaviors persist when more nuclear spin-$\frac{1}{2}$ are included as the LES, only with quantitative modifications of the above quantities.

\begin{figure}
\includegraphics[angle=0]{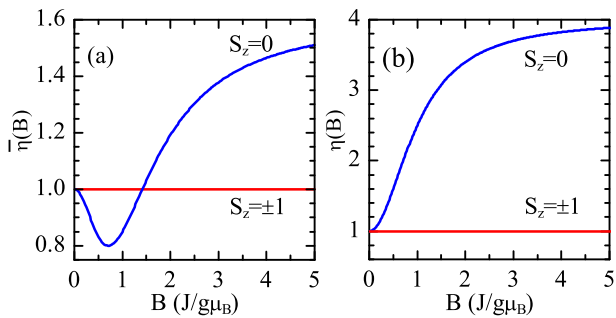}
\caption{(a) The magnetic-field dependence of the time-averaged $\bar{\eta}(B)$ when the state of the composite system at $t=0$ is one of the for four basic states ($|a,b\rangle$, $a,b=\pm\frac{1}{2}$). The results of the $S_z=0$ states show both the negative small-field component and the saturation component at large fields; (b) The magnetic field dependence of  $\eta(B)$ for the eigenstates of the hyperfine-interaction Hamiltonian. They only show the saturation component at large fields, without the ultrasmall-field component.}
\label{Fig4}
\end{figure}

The two-component behavior originates from the dynamical feature of the system. It can be understood analytically by considering four basic states at $t=0$, which are denoted as $|a,b\rangle$ with $a,b=\pm\frac{1}{2}$, where $a$ is for the carrier's spin and $b$ is for the LES. Other states can be seen as their superpositions, hence the results of the four states qualitatively reflect the behavior of Fig. \ref{Fig2}. With the Hamiltonian Eq. (\ref{hami}), the total spin in the $z$ direction, $S_z=I_z+s_z$, is conserved, by which we can classify the four states.  The states $\left|\frac{1}{2},\frac{1}{2}\right\rangle$ ($S_z=1$) and $\left|-\frac{1}{2},-\frac{1}{2}\right\rangle$ ($S_z=-1$) are eigenstates and do not become quantum-correlated states at any time of the evolution with $\eta(t)=1$. However, the other two states $\left|-\frac{1}{2},\frac{1}{2}\right\rangle$ and $\left|\frac{1}{2},-\frac{1}{2}\right\rangle$ ($S_z=0$) become quantum-correlated with time, giving the same result \cite{si}
\begin{equation}
\eta(t)=1-\frac{3}{4}\cdot\frac{\sin^2\omega t+2\alpha^2(1-\cos\omega t)}{(1+\alpha^2)^2},
\end{equation}
where $\omega= J\sqrt{1+\alpha^2}$ and $\alpha\equiv g\mu_BB/J$. A time-independent quantity can be obtained by taking the time-average, which gives
\begin{equation}
\bar{\eta}(B)=1+\frac{3}{5}\cdot\frac{\alpha^2(\alpha^2-2)}{(1+\alpha^2)^2}
\label{average}
\end{equation}
where we have scaled the quantity so that $\bar{\eta}(B=0)=1$. This direct time-average corresponds to the situation when $\nu$ is sufficiently small and the waiting time is much longer than the oscillatory period of $\eta(t)$. Eq. (\ref{average}) is plotted in Fig. \ref{Fig4} (a).
It is clear that $\bar{\eta}$ is less than 1 for $\alpha\le\sqrt{2}$, giving the negative small-field component. The overall behavior can be traced back to the two-fold role of the increasing Zeeman splitting with the magnetic field on the quantum correlation. Take the state $\left|\frac{1}{2},-\frac{1}{2}\right\rangle$ as an example. With vanishing magnetic field, it evolves to the uncorrelated state $\left|-\frac{1}{2},\frac{1}{2}\right\rangle$ halfway in a complete period; with small fields, the degeneracy between the two uncorrelated states is lifted and the system remains correlated before it returns to the initial state; with sufficiently large fields, the energy cost of flipping the carrier's spin allows no sizable probability amplitude of $\left|-\frac{1}{2},\frac{1}{2}\right\rangle$ during the whole process, meaning a negligible degree of quantum correlation and larger hopping ability.

Till now, only the coherent spin dynamics on site $i$ has been considered. However, it should be noted that in some cases the coherency might be disturbed, such as when the spin interactions are enhanced by doping transition metal complex \cite{EpsteinSM}. Although the interaction becomes more complicated, the results based on the hyperfine-interaction model of Eq. (\ref{hami}) are still indicative. In Fig. \ref{Fig4} (b), the magnetic-field-dependent $\eta(B)$ is shown for the eigenstates of the Hamiltonian, which are the states when the composite system reaches thermal equilibrium. Similar to the previous situation, two of the four eigenstates, which are $\left|\frac{1}{2},\frac{1}{2}\right\rangle$ and $\left|-\frac{1}{2},-\frac{1}{2}\right\rangle$ with $S_z=\pm 1$, are uncorrelated states with $\eta(B)=1$. The remaining two eigenstates with $S_z=0$ are correlated ones with \cite{si}
\begin{equation}
\eta(B)=1+\frac{3\alpha^2}{1+\alpha^2}.
\end{equation}
It can be seen that the small-field component is not present, in contrast to the results with coherent spin dynamics. However, the saturation component survives the loss of spin coherence. This implies that the small-field component observed in experiments could be a sign of the spin coherency. The persistence of the saturation component also explains why the OMR is still observed in devices made by hydroxyquinolates consisting of heavy metal atoms \cite{GillinHeavy}.

We further calculate the isotope effect through replacing the spin$-\frac{1}{2}$ protons by the spin$-1$ deuterons. The result is shown Fig. \ref{Fig4}.
\begin{figure}
\includegraphics[angle=0]{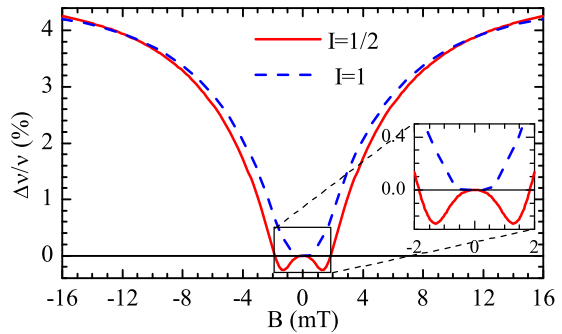}
\caption{The isotope effect obtained by replacing the spin$-\frac{1}{2}$ proton by the spin$-1$ deuteron as the LES. The behavior at small magnetic field is shown in the inset. Compared with the nuclear spin$-\frac{1}{2}$ case, the small-field component for the nuclear spin$-1$ case becomes almost flat.}
\label{Fig5}
\end{figure}
The coupling strength for the nuclear spin$-1$ is chosen to be $0.5J$, where $J$ is that for the nuclear spin$-\frac{1}{2}$. The resulting HWHM of the nuclear spin$-1$ case is larger than that of the nuclear spin$-\frac{1}{2}$ case. Furthermore, only a tiny negative component is present for the nuclear spin-$1$ case, which is shown more clearly in the inset. The difference originates from the intrinsic properties of the system without any further assumptions \cite{si}. The behavior obtained here are in good agreement with the isotope effects from experimental observations \cite{VardenyIsotope}.

\section{Conclusion}

In summary, we have shown that the quenching of the quantum correlation between the carrier's spin and its local environment by the incoherent hopping leads to the OMR. The process contributes an essential prefactor $\eta$ to the attempt-to-escape frequency, which offers a general magnetic-field modulation mechanism. For the hopping of a single carrier to a vacant site with a hyperfine-interaction model, both the saturation component and the negative small-field one emerge naturally. The mechanism holds promise for the incorporation of other influential incoherent processes in the OSC, leading towards a more comprehensive understanding of the magnetic-field effects in these materials.

\textbf{Acknowledgments}

The authors would like to thank V. Dediu, A. J. Drew and W. P. Gillin for their helpful discussions and M. Willis for the reading of this manuscript. We acknowledge the financial supports from the National Natural Science Foundation of China and the National Basic Research Program  of China (2012CB921401 and 2009CB929204).

\end{document}